\documentclass{svproc}
\usepackage{url}

\usepackage{graphicx}
\usepackage{bigints}

\newcommand{\MeV}{\, {\rm MeV}}

\begin{document}
\mainmatter              % start of a contribution
\title{Cross-correlators of conserved charges in QCD}
\titlerunning{}  % abbreviated title (for running head)

\author{R. Bellwied\inst{1}, S. Borsanyi\inst{2}, Z. Fodor\inst{2,3,4,5}, J. N. Guenther\inst{2,6}, J. Noronha-Hostler\inst{7}, P.Parotto \inst{1,2}, A. P\'asztor\inst{3}, C. Ratti\inst{1}, J. M. Stafford\inst{1}}
\authorrunning{R. Bellwied \textit{et al.}} % abbreviated author list (for running head)
\tocauthor{}
\institute{Department of Physics, University of Houston, Houston, TX, USA  77204 
%\email{I.Ekeland@princeton.edu},\\ WWW home page:
%\texttt{http://users/\homedir iekeland/web/welcome.html}
\and
University of Wuppertal, Department of Physics, Wuppertal D-42119, Germany
\and 
E{\"o}tv{\"o}s University, Budapest 1117, Hungary
\and 
J{\"u}lich Supercomputing Centre, J{\"u}lich D-52425, Germany
\and
UCSD, Physics Department, San Diego, CA 92093, USA
\and
University of Regensburg, Department of Physics, Regensburg D-93053, Germany
\and
Department of Physics, University of Illinois at Urbana-Champaign, Urbana, IL 61801, USA
}
\maketitle              % typeset the title of the contribution
\begin{abstract}
We present cross-correlators of QCD conserved charges at $\mu_B=0$ from lattice simulations and perform a Hadron Resonance Gas (HRG) model analysis to break down the hadronic contributions to these correlators. We construct a suitable hadronic proxy for the ratio $-\chi_{11}^{BS}/\chi_2^S$  and discuss the dependence on the chemical potential and experimental cuts. We then perform a comparison to preliminary STAR results and comment on a possible direct comparison of lattice and experiment.
% We would like to encourage you to list your keywords within
% the abstract section using the \keywords{...} command.
\end{abstract}
\section*{Introduction and setup}

The transition between hadronic matter and deconfined Quark Gluon Plasma (QGP) is a
smooth crossover at vanishing baryon chemical potential \cite{Aoki:2006we,Borsanyi:2010bp,Bazavov:2011nk,Bazavov:2018mes}, and is believed to turn into a first order 
transition for larger values of the chemical potential. Among the best suited observables 
for the study of the QCD phase diagram in the transition region are fluctuations of conserved charges. They can be studied in theory through first principles lattice QCD 
calculations (see e.g. \cite{Borsanyi:2011sw,Bazavov:2012jq,Bazavov:2012vg,Borsanyi:2013hza,DElia:2016jqh}), as well as being closely connected to experimentally available 
measurements of net-particle fluctuations and correlations \cite{Adamczyk:2013dal,Adamczyk:2014fia,Adamczyk:2017wsl,Nonaka:2019fhk,Adam:2019xmk}. 
Due to the fact that some hadrons cannot be detected in  
experiment, a sizable share of $B,Q,S$ is lost. Historically, the hadronic proxies used for 
$B$, $Q$ and $S$ are protons, the sum $p+\pi+K$ and the kaons themselves, 
respectively. More recently, the attention has moved towards non-diagonal correlators 
between conserved charges \cite{Nonaka:2019fhk,Adam:2019xmk}. 

In this contribution we build a bridge between lattice-QCD-calculated correlators of 
conserved charges and experimentally accessible fluctuations and correlations of 
hadronic species, focusing on the correlator between baryon number and strangeness $\chi_{11}^{BS}$. We employ the HRG model in order to include 
the effect of resonance decays and cuts on the kinematics, which are present in the 
experiment. Most 
importantly, the HRG model allows us to isolate single particle-particle correlations and 
connect them to the correlators of conserved charges.

The ideal HRG model partition function is a sum over the single-state partition functions.
Fluctuations of conserved charges are expressed as derivatives of the grand partition 
function with respect to the different chemical potentials:
\begin{equation} 
\chi^{BQS}_{ijk}(T, \hat{\mu}_B, \hat{\mu}_Q, \hat{\mu}_S) = \frac{\partial^{i+j+k} \left( p/T^4 \right)}{\partial \hat{\mu}_B^i \partial \hat{\mu}_Q^j \partial \hat{\mu}_S^k} =\sum_R B_R^i \, Q_R^j \, S_R^k \,  I^R_{i+j+k} \left(T, \hat{\mu}_B, \hat{\mu}_Q, \hat{\mu}_S \right) \, \, ,
\label{eq:chiBQS}
\end{equation}
where $\hat{\mu_i} = \mu_i/T$, and the phase space integral at order $i+j+k$ reads (note 
that it is completely symmetric in all indices, hence $i+j+k = l$):
\begin{equation}
I^R_{l} \left(T, \hat{\mu}_B, \hat{\mu}_Q, \hat{\mu}_S \right) = \frac{\partial^{l} p_R/T^4}{\partial \hat{\mu}_R^{l}} \, \, .
\end{equation}

\vspace{-6mm}
\begin{figure}[h]
\vspace{-9mm}
\begin{minipage}{0.45\textwidth}
\includegraphics[width=1.0\textwidth]{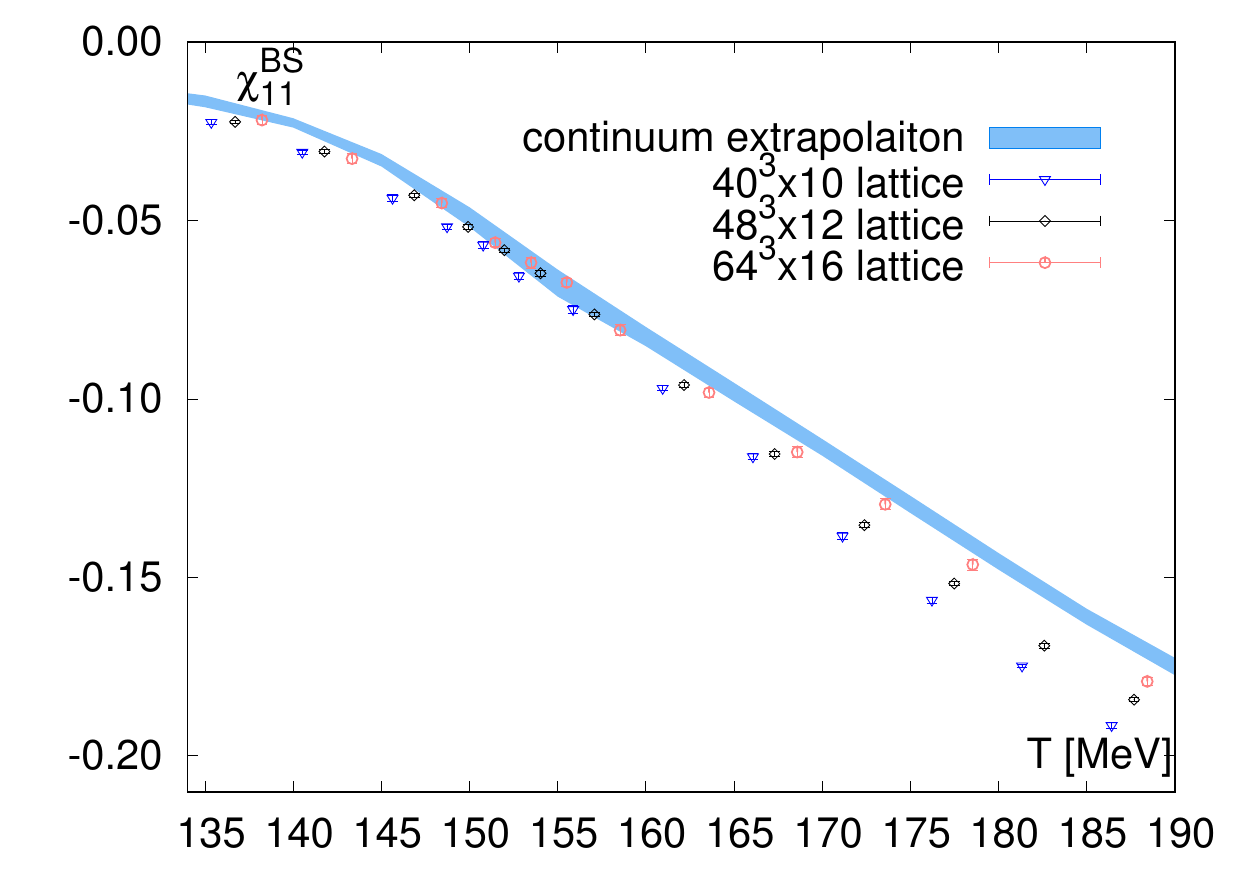}
\end{minipage}
\begin{minipage}{0.5\textwidth}
\vspace{6mm}
\includegraphics[width=0.97\textwidth]{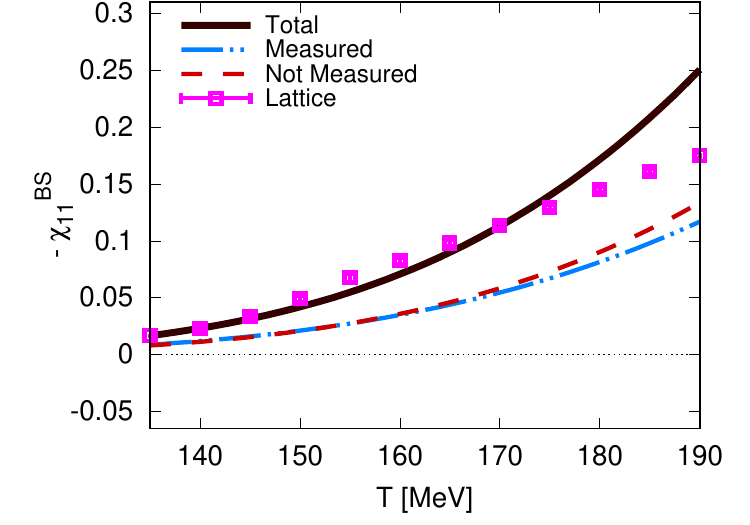}
\end{minipage} \vspace{-4mm}
\caption{The correlator $\chi_{11}^{BS}$ as a function of the temperature at 
$\mu_B=0$. (Left panel) results for the $\omega_0$-based scale setting \cite{Borsanyi:2012zs} at finite lattice spacing and continuum extrapolation. (Right panel) the
continuum extrapolated results compared to HRG model calculations (solid black line),
with the contribution from measured (dotted-dashed blue line) and
non-measured (dashed reline) hadronic species. Figure from \cite{Bellwied:2019pxh}.}\label{fig:BS_lat}
\end{figure}

\vspace{-3mm}
It is possible to recast the sum in Eq. \ref{eq:chiBQS} as a sum over the fewer states 
which are stable under strong interactions:
\begin{equation}
\sum_R B_R^l Q_R^m S_R^n I^R_p \rightarrow \sum_{i \in \rm stable} \sum_R \left( P_{R \rightarrow i} \right)^p B^l_i Q^m_i S^n_i I^R_p\, \, ,
\label{eq:daughters}
\end{equation}
where $\left( P_{R \rightarrow i} \right)^p$ is the average number of particles $i$ 
produced by the decay of particle $R$.

The advantage of expressing the fluctuations in Eq. \ref{eq:chiBQS} in term of stable 
particles, is that we can further distinguish by particles which can be -- or usually are -- 
detected in experiment, and those which are not. In this work we employ the hadronic list labeled as PDG2016+ in \cite{Alba:2017mqu}, with the list of decays described and first utilized in \cite{Alba:2017hhe}. We will hereafter consider the 
following species as the commonly measured ones:
$$\pi^\pm , \, \,  K^\pm , \, \,  p \left( \overline{p}\right) , \, \,  \Lambda ( \overline{\Lambda}) , \, \,  \Xi^- ( \overline{\Xi}^+) , \, \,  \Omega^- ( \overline{\Omega}^+) .$$

\begin{figure}[h!]
\includegraphics[width=0.5\textwidth]{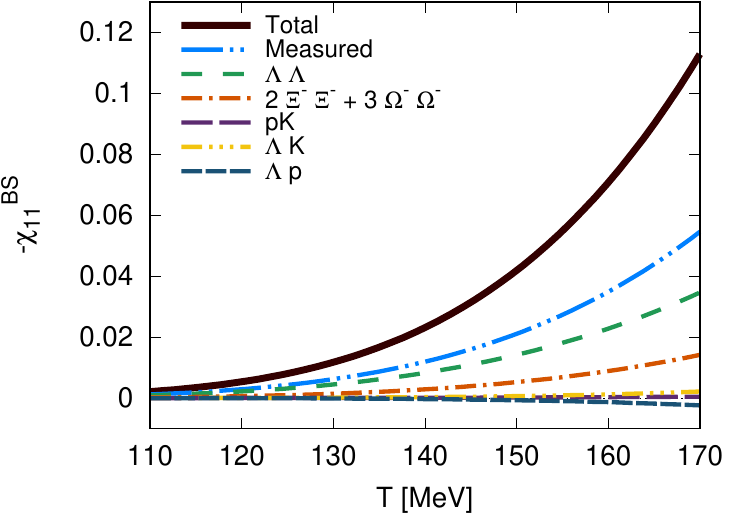}
\includegraphics[width=0.5\textwidth]{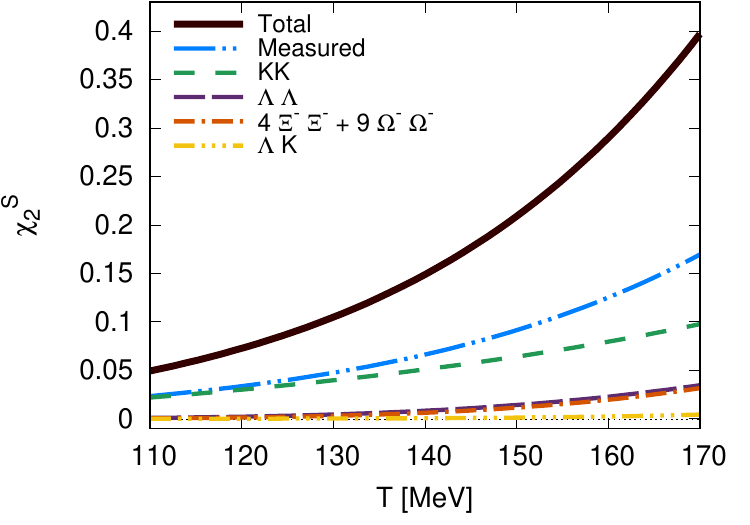}
\caption{Breakdown of the contributions from \textit{measured} particle-particle correlators to $-\chi_{11}^{BS}$ (left panel) and $\chi_2^S$ (right panel), at $\mu_B=0$, as a function of the temperature. Figure from \cite{Bellwied:2019pxh}. \vspace{-5mm}} \label{fig:brkdwn}
\end{figure}
\vspace{-1mm}

In Fig. \ref{fig:BS_lat} we show the $\chi_{11}^{BS}$ correlator as a function of the
temperature for vanishing chemical potential, calculated from the lattice (left panel) at 
different finite spacings, as well as its continuum extrapolation. In the right panel, along
with the continuum extrapolation, we show the results from our HRG model analysis,
where we separate the contribution to this correlator from measured and non-measured
hadronic species. We see that the contributions roughly correspond to the same amount.
Moreover, we can see in Fig. \ref{fig:brkdwn} the breakdown of the main contributions
from \textit{measured} particle-particle correlations to $-\chi_{11}^{BS}$ and $\chi_2^S$
at vanishing chemical potential.

\section*{Proxy for $-\chi_{11}^{BS}/\chi_2^S$}

\begin{figure}[h]
\includegraphics[width=0.46\textwidth]{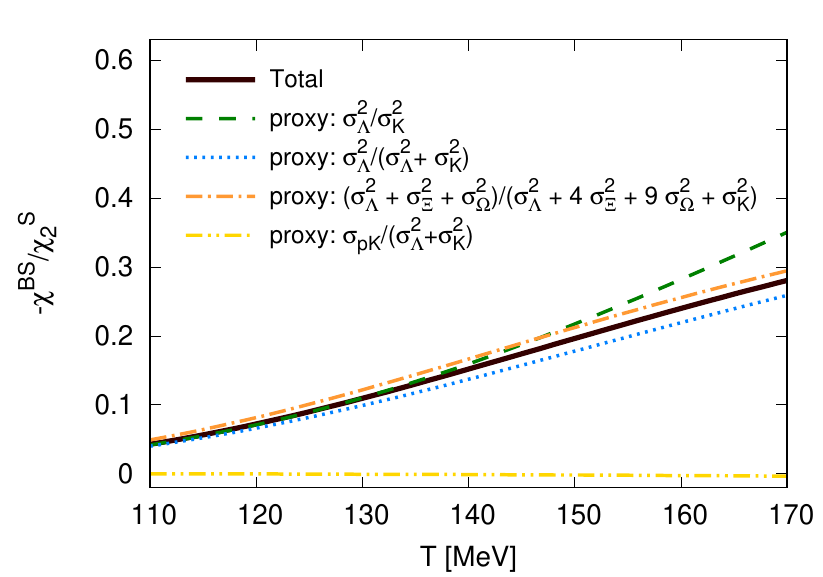}
\includegraphics[width=0.43\textwidth]{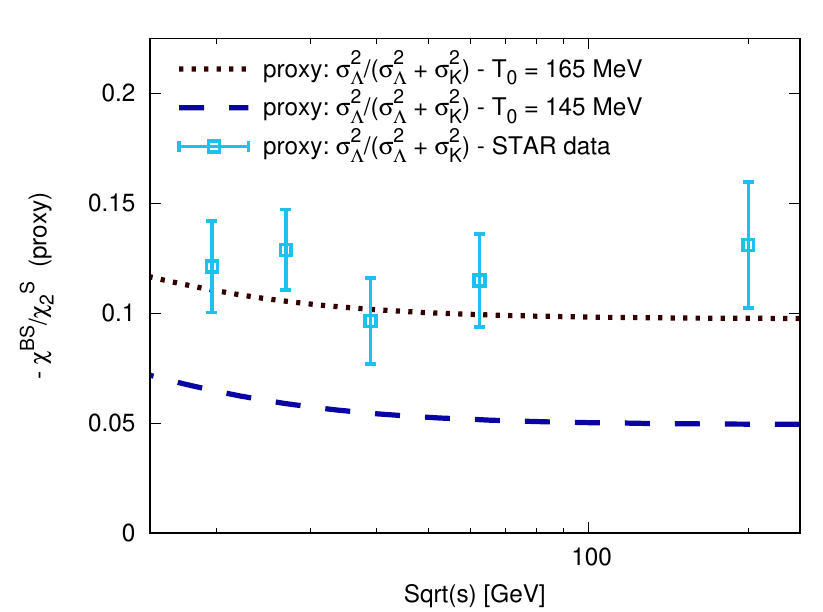}
\caption{(Left panel): comparison of different proxies (the proxy $\widetilde{C}^{\Lambda,\Lambda K}_{BS,SS}$ is shown as a blue dotted line) and the total contribution (black solid line) for the ratio $-\chi_{11}^{BS}/\chi_2^S$ at $\mu_B=0$. (Right panel): comparison of our proxy with the kinematic cuts from \cite{Adamczyk:2017wsl,Nonaka:2019fhk}, along parametrized freeze-out lines with $T(\mu_B=0) = 145, 165 \MeV$ (black dotted and blue dashed). The STAR preliminary data are shown in light blue. Figure from \cite{Bellwied:2019pxh}.} \label{fig:pxy_muB0_exp}
\end{figure}

In order to perform a comparison to experiment and potentially to lattice QCD results,
we consider the ratio $-\chi_{11}^{BS}/\chi_2^S$. Exploiting the information in Fig. \ref{fig:brkdwn}, we construct the following proxy for this ratio:
\begin{equation}\label{eq:pxy_BS_SS_2}
\widetilde{C}^{\Lambda,\Lambda K}_{BS,SS} = \sigma_\Lambda^2/( \sigma_K^2 + \sigma_\Lambda^2) \, \, ,
\end{equation} 
which is shown in the left panel of Fig. \ref{fig:pxy_muB0_exp} as a blue dotted line,
alongside the ratio $-\chi_{11}^{BS}/\chi_2^S$ (black solid line). This quantity
well reproduces the full contribution for all temperatures around the QCD transition.

\begin{figure}
\includegraphics[width=\textwidth]{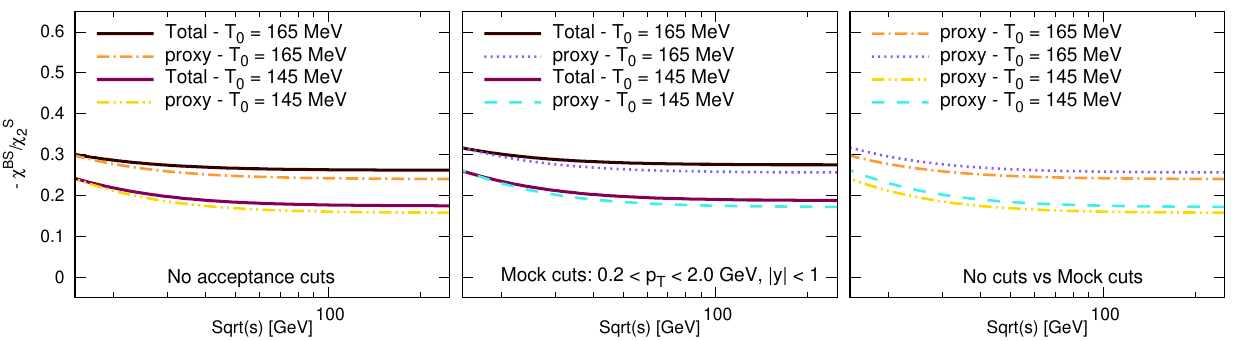}
\caption{Comparison of the proxy $\widetilde{C}^{\Lambda,\Lambda K}_{BS,SS}$ to the ratio $-\chi_{11}^{BS}/\chi_2^S$ for finite chemical potential, along the same freeze-out lines as in Fig. \ref{fig:pxy_muB0_exp}. The left panel shows the comparison without cuts, while the central panel with ``exemplary'' cuts. In the right panel the proxy is shown both with and without cuts. Figure from \cite{Bellwied:2019pxh}. \vspace{-6mm}}\label{fig:pxy_allcuts}
\end{figure}

In Fig. \ref{fig:pxy_allcuts} we show the same comparison for finite chemical potential, along parametrized chemical freeze-out lines with $T(\mu_B=0) = 145, 165 \MeV$. In the left panel we show the comparison in the absence of cuts on the kinematics, while in the central panel we introduce ``exemplary'' cuts, which are the same for all the hadronic species; in both cases we see that the proxy works well for a broad range of collision energies. In the right panel we compare our proxy in the case with and without the cuts, and notice that the effect of the cuts is quite modest. This hints at the possibility of directly comparing lattice QCD calculations and experimental measurements for this particular ratio.

In the right panel of Fig.\ref{fig:pxy_muB0_exp} we compare preliminary STAR results to our proxy, where we have utilized the same cuts as present in the experimental analysis \cite{Adamczyk:2017wsl,Nonaka:2019fhk}. We see that a higher chemical freeze-out temperature is preferred, which is in line with previous findings \cite{Bellwied:2018tkc,Bluhm:2018aei}. A direct comparison to lattice QCD results is however premature, since it is essential that the same cuts are applied to all hadronic species for the proxy we constructed to reproduce the ratio $-\chi_{11}^{BS}/\chi_2^S$.

\section*{Acknowledgements}
This project was partly funded by the DFG grant SFB/TR55 and also
supported by the Hungarian National Research,  Development and
Innovation Office, NKFIH grants KKP126769 and K113034. The project
also received support from the BMBF grant 05P18PXFCA.
Parts of this work were supported  by  the National Science Foundation  under
grant  no.  PHY-1654219 and by the U.S.  Department of
Energy,  Office  of Science,  Office  of  Nuclear Physics, within the framework
of the Beam Energy Scan Theory (BEST) Topical Collaboration. 
A.P. is supported by the J\'anos Bolyai Research Scholarship of the
Hungarian Academy of Sciences and by the \'UNKP-19-4 New National Excellence
Program of the Ministry of Innovation and Technology.
The authors gratefully acknowledge the Gauss Centre for Supercomputing e.V.
(www.gauss-centre.eu) for funding this project by providing computing time on
the GCS Supercomputer JUWELS and JURECA/Booster at J\"ulich Supercomputing
Centre (JSC), and on SUPERMUC-NG at LRZ, Munich
 as well as on HAZELHEN at HLRS Stuttgart, Germany.  C.R. also
acknowledges the support from the Center of Advanced Computing and Data Systems
at the University of Houston. J.N.H. acknowledges the support of the Alfred P.
Sloan Foundation, support from the US-DOE Nuclear Science Grant No.
de-sc0019175. R.B. acknowledges support from the US DOE Nuclear Physics Grant
No. DE-FG02-07ER41521.

%\section*{Conclusions}
%
%In this contribution we have shown a strategy to perform a direct comparison between
%lattice-QCD-calculated correlators on conserved charges $B,Q,S$ and a suitable ratio experimentally measured moments of net-particle distributions.


\end{document}